\documentclass[fp,twocolumn]{jpsj3}
\def\etal.{et\penalty50\ al.}
\usepackage{txfonts}
\usepackage{graphicx,bm,color,braket}
\usepackage[mode=text]{siunitx}

\title{Zero-Field Surface Charge Due to the Gap Suppression in $d$-Wave Superconductors}

\author{Ezekiel Sambo Joshua$^1$, Hikaru Ueki$^2$, Wataru Kohno$^1$, and Takafumi Kita$^1$}
\inst{
$^1$Department of Physics, Hokkaido University, Sapporo 060-0810, Japan\\
$^2$Department of Mathematics and Physics, Hirosaki University, Hirosaki, Aomori 036-8561, Japan
} 

\abst{
We perform a microscopic study on the redistribution of electric charge near the surface of a model $d$-wave superconductor cut along the [110] direction,
with a Fermi surface appropriate for cuprate superconductors, using the augmented quasiclassical equations.
We identify two possible mechanisms for the redistribution of charged particles different from the well-known magnetic Hall effect, namely; the pair potential gradient (PPG) force due to surface effects on the pair potential and the pressure difference between the normal and superconducting regions arising from the slope of the density of states (SDOS) in the normal states at the Fermi level. 
Our present results show that in spite of the absence of supercurrents, electric charge is induced around the surface.
Moreover, the charging effect due to the SDOS pressure dominates over that due to the PPG force 
for all the realistic electron-fillings $n=0.8$, $0.9$, and $1.15$, at all temperatures. In addition, for the filling  $n=1.15$, the PPG force and the SDOS pressure contributions have the same negative signs, which gives a larger total surface charge
i.e., both the sign and amount of the surface charge depends greatly on the Fermi-surface curvature. 
We have also calculated the local density of states (LDOS) within the augmented quasiclassical theory. Spatially varying local particle-hole asymmetry appears in the LDOS, which suggests the presence of electric charge. 
}

\begin{document}
\maketitle

\section{Introduction \label{sec:I}} 
In general, electrostatic charge redistribution implies the action of certain forces on charged particles. For instance, in normal metals\cite{Hall}, semiconductors\cite{Prange} as well as in superconductors\cite{London,Kita01,Ueki16}, the magnetic Lorentz force results in the Hall effect. Exclusively in superconductors, in the presence of inhomogeneities such as surfaces or interfaces,  other forces  appear notwithstanding the absence of magnetic fields, namely; the pair-potential gradient (PPG) force\cite{Kopnin94,Arahata14,Ohuchi} which originates from the spatial variation of the pair potential, and the pressure due to the slope of the density of states (SDOS) in the normal states at the Fermi level\cite{Khomskii92,Khomskii95,Ueki18}. These forces appear in the vortex state in type-II  superconductors and are expected in the presence of surfaces and interfaces.

Quite recently, the augmented quasiclassical equations incorporating the three force terms were derived, with the standard Eilenberger equations\cite{Eilenberger,KitaText} as the leading-order contributions
 and the force terms as first-order corrections in terms of the quasiclassical parameter 
 $\delta\equiv\hbar/\langle p_{\rm F} \rangle_{\rm F}\xi_0\ll1$,
 where $p_{\rm F}$ is the magnitude of the Fermi momentum, 
 $\langle \cdots \rangle_{\rm F}$ is the Fermi surface average, 
 and $\xi_0$ is the coherence length at zero temperature\cite{Kita01,Arahata14,Ueki18}.
 These augmented quasiclassical equations have been used in the study of the Hall effect in superconductors in the Meissner state\cite{Kita09},
 and in the vortex state\cite{Arahata14,Ohuchi,Ueki16,Ueki18,Masaki19,Kohno}. However, to the best of our knowledge, they have not yet been applied to surface systems.

It has been shown that in the isolated vortex core of an isotropic type-II superconductor the PPG force gives up to $10$ to $10^2$ times larger contribution to the electric charging compared to the  Lorentz force\cite{Ohuchi}. Even more recently, Ueki {\it et al.} found that the SDOS pressure gives the dominant contribution near the transition temperature, while the PPG force dominates as the temperature approaches zero\cite{Ueki18}. Masaki studied the charged and uncharged vortices in a chiral $p$-wave superconductor based on the augmented quasiclassical equations. He pointed out that the vortex-core charge is dominated by the contributions from the angular derivative terms in the PPG force terms\cite{Masaki19}. Using a simplified picture of a system consisting of a vortex core in the normal state surrounded by a superconducting material, Khomskii {\it et al.}
showed that a finite difference in chemical potential $\delta\mu\neq0$ between the normal and the superconducting subsystems results in the redistribution of charge\cite{Khomskii92,Khomskii95}. 
In the context of surface charging, Furusaki {\it et al.} studied spontaneous surface charging in chiral $p$-wave superconductors based on the Bogoliubov--de Gennes (BdG) equations. They found two contributions, one contribution originates from the Lorentz force due to the spontaneous edge currents, while the other contribution has topological origin and is related to the intrinsic angular momentum of the Cooper pairs\cite{Furusaki}. They also calculated the surface charging due to the Lorentz force acting on the spontaneous edge currents in $d+is$-wave superconductors based on the BdG equations\cite{Furusaki}. Emig {\it et al.} also showed using a phenomenological analysis on the basis of Ginzburg--Landau theory that the presence of surfaces in $d$-wave superconductors can induce charge inhomogeneity due to the suppression of the energy gap\cite{Emig}.
Volovik and Salomaa discussed the appearance of electric charge at the edges and vortex core of even electrically neutral $p$-wave superfluids\cite{Volovik,Salomaa}.

In a $d_{x^2-y^2}$-wave superconductor with a specularly reflective surface cut along the [110] direction, 
the  order parameter is suppressed \cite{Buchholtz95,Nagato95,Yan98} near the surface and vanishes at the surface,
due to a change in its sign along the classical quasiparticle trajectories. 
This sign change also results in the formation of zero energy states (ZES) near (at) the edges of these materials\cite{Tanuma98,Fogelstrom98,Hu94,Tanaka,Yang,Kashiwaya,Lofwander,Yada,Sato}. 
ZES in $d$-wave superconductors are detectable through the observation of 
zero-bias conductance peaks in the spectra of scanning tunneling spectroscopy at oriented surfaces of the $d$-wave crystals\cite{Geerk,Lesueur,Covington}.
Hayashi {\it et al.} discussed the connection between the Caroli--de Gennes--Matricon states\cite{Caroli} at the vortex core of an $s$-wave superconductor and the occurrence of electric charge\cite{Hayashi}. They concluded that the particle-hole asymmetry inside the vortex core observed through the local density of states (LDOS) implies the corresponding existence of charge at the vortex core. Recently, Masaki also discussed the connection between particle-hole asymmetry and vortex charging in superconductors\cite{Masaki19}. Surface charging in $d$-wave superconductors may also have a similar connection with particle-hole asymmetry in the LDOS, which is expected to appear due to first-order quantum corrections within the augmented quasiclassical theory.

Even in the absence of magnetic fields, surface effects in $d$-wave superconductors
result in the appearance of the PPG force
due to the suppression of the pair potential near the surface and the pressure
due to the SDOS at the Fermi level. 
Hence the presence of oriented surfaces in $d$-wave superconductors is expected 
to be accompanied by the redistribution of charged particles.

In this paper, we report the accumulation of charged particles around the [110] specular surface of a $d_{x^2-y^2}$-wave superconductor with a Fermi surface used for cuprate superconductors, 
due to the PPG force and the SDOS pressure,  notwithstanding the absence of magnetic field.

Although surface charging without time-reversal symmetry breaking itself has already been suggested in Ref.\ \citen{Emig}, our microscopic theory clarifies the origin of the charging, which is missing from the Eilenberger equations. 
The augmented quasiclassical theory describes the particle-hole-asymmetric LDOS which also cannot be described at the level of the Eilenberger equations.
In our calculations, it is shown that
the SDOS pressure gives the dominant contribution to the 
surface charge for the realistic electron-fillings $n=0.8$, $0.9$, and $1.15$ at all temperatures.
Moreover, the SDOS pressure with some approximations reduces to the mechanism considered in Ref.\ \citen{Emig}. 
It is also noteworthy that since the PPG-force contribution turns out to be comparable to that due to the SDOS pressure, their respective contributions should be calculated simultaneously. 
We also observe ZES at the surface of the $d$-wave superconductor within the augmented quasiclassical theory. In addition, we discuss the structure of the LDOS, especially the particle-hole asymmetry due to the SDOS pressure and the PPG force as an indicator of the existence of surface charge. This asymmetric behaviour in the LDOS caused by surface charging is expected to be observed.

This paper is organized as follows. In Sect.\ \ref{sec:II}, we give a summary of the quasiclassical formalism relevant to our present study. This is based on the formulation in Ref. \citen{Ueki18}. Our numerical procedures  and  results are summarized in Sect.\ \ref{sec:III}.  In Sect.\ \ref{sec:IV}, we give a summary of the present study and discuss future directions.

\section{Formalism \label{sec:II}}
\subsection{Augmented Eilenberger equations \label{subsec:AI}}
The quasiclassical equations  augmented with the PPG terms are given in the Matsubara formalism by \cite{Ohuchi,Ueki18}
\begin{align}
&\left[ i\varepsilon_n \hat{\tau}_3 - \hat{\Delta} \hat{\tau}_3, \hat{g} \right] 
+ i \hbar {\bm v}_{\rm F} \cdot {\bm \partial} \hat{g} \notag \\
&\ \ \ - \frac{i\hbar}{2} 
\left\{{\bm\partial}\hat{\Delta}\hat{\tau}_3, {\bm\partial}_{{\bm p}_{\rm F}}\hat{g}\right\}
+ \frac{i\hbar}{2} 
\left\{{\bm\partial}_{{\bm p}_{\rm F}}\hat{\Delta}\hat{\tau}_3, {\bm\partial}\hat{g}\right\} 
= \hat{0}, \label{AEEq}
\end{align}
where $\varepsilon_n=(2n+1)\pi k_{\rm B}T$ is the fermion Matsubara energy $(n=0,\pm1,\dots)$, 
${\bm v}_{\rm F}$ is the Fermi velocity
while ${\bm p}_{\rm F}$ is the Fermi momentum,
and $\bm \partial$ is the gauge-invariant differential operator \cite{Ohuchi,Ueki18}.
The commutators are given by
$[\hat{a},\hat{b}]\equiv{\hat{a}\hat{b}-\hat{b}\hat{a}}$ and $\{\hat{a},\hat{b}\}\equiv{\hat{a}\hat{b}+\hat{b}\hat{a}}$.
The functions 
$\hat{g}=\hat{g}(\varepsilon_n,\bm p_{\rm F},\bm r)$ 
and $\hat{\Delta} = \hat{\Delta} ({\bm p}_{\rm F}, {\bm r})$ 
are the quasiclassical Green's functions and the pair potential, respectively. 
We have used the static gauge as ${\bm E} ({\bm r}) = - {\bm \nabla} \Phi ({\bm r})$ 
and ${\bm B} ({\bm r}) = {\bm \nabla} \times {\bm A} ({\bm r})$, 
where ${\bm E}$ is the electric field, ${\bm B}$ is the magnetic field, $\Phi$ is the scalar potential, 
and ${\bm A}$ is the vector potential.

We consider the spin-singlet pairing state without spin paramagnetism. 
The matrices $\hat{g}$, $\hat{\Delta}$ and $\hat{\tau}_3$ are written as
\begin{align}
\hat{g}=
\begin{bmatrix}
{g} & - i f \\
i \bar{f} & - \bar{g}
\end{bmatrix}, \ \ \ 
\hat{\Delta}=
\begin{bmatrix}
{0} & \Delta \phi \\
\Delta^* \phi & {0}
\end{bmatrix},\ \ \ 
\hat{\tau}_3 =
\begin{bmatrix}
1 & 0 \\
0 & -1
\end{bmatrix},
\end{align}
where the functions $\Delta = \Delta ({\bm r})$ and $\phi = \phi ({\bm p}_{\rm F})$ 
denote the amplitude of the energy gap and the basis function of the energy gap, respectively. 
The barred functions in the Matsubara formalism are defined generally by 
$\bar{X} (\varepsilon_n, {\bm p}_{\rm F}, {\bm r}) 
\equiv X^* (\varepsilon_n, - {\bm p}_{\rm F}, {\bm r})$.

Following the procedure used  in Ref. \citen{Kita09}, we carry out a perturbation expansion of $f$ and $g$  in terms of $\delta$ 
as $f=f_0+f_1+\cdots$ and $g=g_0+g_1+\cdots$.
Then the main part in the standard Eilenberger equations \cite{KitaText,Eilenberger} 
are obtained from the leading-order contribution in terms of the quasiclassical parameter as 
\begin{align}
\varepsilon_nf_0
+\frac{1}{2}\hbar{\bm v}_{\rm F}\cdot \left({\bm \nabla} - i \frac{2e{\bm A}}{\hbar}\right) {f}_0
=\Delta{\phi}{g}_0, 
\label{Eeq}
\end{align}
where $e < 0$ is the electron charge.
We also obtain the relations from the normalization condition as 
$g_0 = \bar{g}_0 = {\rm sgn}(\varepsilon_n) \sqrt{1-f_0\bar{f}_0}$ \cite{KitaText}. 
Moreover, the self-consistency equation for the pair potential is given by \cite{KitaText} 
\begin{align}
\Delta = 
2\pi\Gamma k_{\rm{B}}{T}\sum_{n=0}^{n_{\rm c}} 
 \langle{f_{0} \phi}\rangle_{\rm F}, 
\label{gapeq}
\end{align}
where the cutoff $n_{\rm c}$ is determined from $(2n_{\rm c}+1)\pi k_{\rm B}T = \varepsilon_{\rm c}$ with $\varepsilon_{\rm c}$ denoting the cutoff energy\cite{KitaText}, and
$\Gamma$ denotes the coupling constant responsible for the  Cooper pairing, defined by
$\Gamma \equiv - N(0) V_{\rm eff}$ with $V_{\rm eff}$ and $N(0)$ denoting 
the constant effective potential 
and the normal-state density-of-states (DOS) per spin and unit volume at the Fermi level, respectively. 
We obtain the expression for the first-order Green's function $g_1$ in terms of quasiclassical parameter $\delta$ from Eq. (\ref{AEEq}) as
\begin{align}
&{\bm v_{\rm F}} \cdot {\bm \nabla} g_1 \notag \\
& \ \ \ = -\frac{i}{2} \left({\bm \nabla} + i \frac{2e{\bm A}}{\hbar}\right) \Delta^* \phi 
\cdot \frac{\partial {f_0}}{\partial {\bm p_{\rm F}}} 
- \frac{i}{2} \left({\bm \nabla} - i \frac{2e{\bm A}}{\hbar}\right) \Delta \phi 
\cdot \frac{\partial {\bar{f_0}}}{\partial {\bm p_{\rm F}}} \notag \\
& \ \ \ + \frac{i}{2} \Delta^* \frac{\partial \phi}{\partial {\bm p_{\rm F}}} 
\cdot \left({\bm \nabla} - i \frac{2e{\bm A}}{\hbar}\right) f_0
+ \frac{i}{2} \Delta \frac{\partial \phi}{\partial {\bm p_{\rm F}}} 
\cdot \left({\bm \nabla} + i \frac{2e{\bm A}}{\hbar}\right) {\bar{f}_0}.\label{g1Eq}
\end{align}
We note that the momentum derivative of $\phi$ terms come in the $g_1$ equation for anisotropic superconductors\cite{Masaki19}.

\subsection{Local density of states}
The LDOS is obtained as \cite{Ueki18} 
\begin{align}
&{{N_{\rm s}}}(\varepsilon,\bm r) = 
N(0) \langle {\rm Re} g_0^{\rm R} + {\rm Re} g_1^{\rm R} \rangle_{\rm F} 
+ N'(0) \varepsilon \langle {\rm Re} g_0^{\rm R} \rangle_{\rm F} \notag \\
& \ \ \ + \frac{N'(0)}{2} \langle {\rm Im} (f_0^{\rm R} \Delta^* \phi) + {\rm Im} (\bar{f}_0^{\rm R} \Delta \phi) \rangle_{\rm F},\label{LDOS}
\end{align}
where the functions 
$g_{0,1}^{\rm R}$ and $f_0^{\rm R}$ are the quasiclassical retarded Green's functions 
which are obtained by solving Eqs. (\ref{Eeq}) and (\ref{g1Eq}) with the following transformation:
$g_{0,1}^{\rm R} (\varepsilon, {\bm p}_{\rm F}, {\bm r}) = g_{0,1} (\varepsilon_n\to{-i\varepsilon+\eta}, {\bm p}_{\rm F}, {\bm r})$ 
and 
$f_0^{\rm R} (\varepsilon, {\bm p}_{\rm F}, {\bm r}) = f_0 (\varepsilon_n\to{-i\varepsilon+\eta}, {\bm p}_{\rm F}, {\bm r})$, 
$\eta$ is a positive infinitesimal constant, 
and the barred functions in the real energy formalism are defined generally by 
$\bar{X} (\varepsilon, {\bm p}_{\rm F}, {\bm r}) 
\equiv X^* (- \varepsilon, - {\bm p}_{\rm F}, {\bm r})$.

\subsection{Electric field equation}
The electric field is expressed as\cite{Ueki18}
\begin{align}
&-\lambda^2_{\rm TF}{\bm \nabla}^2{\bm E}(\bm r)+{\bm E}(\bm r)=-\frac{2\pi{k_{\rm{B}}T}}{e}\sum_{n=0}^{\infty}\left\langle{{\bm \nabla}{{\rm Im}g_1}}\right\rangle_{\rm F} \notag \\
& \ \ \  -\frac{1}{e} \frac{N'(0)}{N(0)}\int_{\tilde{\varepsilon}_{{\rm c}-}}^{\tilde{\varepsilon}_{{\rm c}+}}d\varepsilon \bar{n} (\varepsilon){\varepsilon}\left\langle{\bm \nabla}{\rm Re}g_0^{\rm R} \right\rangle_{\rm F}
 {-\frac{c}{e}}\frac{N'(0)}{N(0)}{\bm \nabla} {|\Delta|^2}, \label{AQCEq} 
\end{align}
where $\lambda_{\rm TF}\equiv\sqrt{d \epsilon_0/2e^2N(0)}$ denotes the Thomas\textendash Fermi screening length with thickness $d$ \cite{Masaki19,Artemenko} and the vacuum permittivity $\epsilon_0$, 
and the function $\bar{n} (\varepsilon)={1}/{({\rm e}^{\varepsilon/k_{\rm B}T}+1})$ is the Fermi distribution function for quasiparticles.
The first term on the RHS of Eq. (\ref{AQCEq}) is the PPG term, while the second and third terms are the contributions from the SDOS pressure. Furthermore, it can be seen that the third term depends on the gradient of the amplitude of the pair potential. 
The parameter $c$ first introduced by Khomskii {\it et al.}\cite{Khomskii95} is given by\cite{Ueki18}
\begin{align}
c\equiv\int_{\tilde{\varepsilon}_{{\rm c}-}}^{\tilde{\varepsilon}_{{\rm c}+}}d\varepsilon\frac{1}{2\varepsilon}\tanh\frac{\varepsilon}{2k_{\rm{B}}T_{\rm{c}}},
\end{align}
 where $T_{\rm{c}}$ denotes the superconducting transition temperature at zero magnetic field.
The cutoff energies $\tilde{\varepsilon}_{{\rm c}\pm}$ are determined by\cite{Ueki18} 
\begin{align}
\int_{\tilde{\varepsilon}_{{\rm c}-}}^{\tilde{\varepsilon}_{{\rm c}+}}N_{\rm s}(\varepsilon,\bm r)d\varepsilon
=\int_{\tilde{\varepsilon}_{{\rm c}-}}^{\tilde{\varepsilon}_{{\rm c}+}}N(\varepsilon)d\varepsilon,
 \ \ \ N_{\rm s}(\tilde{\varepsilon}_{{\rm c}\pm},\bm r) = N (\tilde{\varepsilon}_{{\rm c}\pm}). \label{cutoffenergyeq}
\end{align}
%

\subsection{Density of states and the chemical potential difference in the homogeneous system}
We introduce the normal DOS, $N({\varepsilon})$, expressed as

\begin{align}
N(\varepsilon)\equiv
\int_{1^{\rm st}{\rm BZ}} \frac{d{p}_{x}d{p}_{y}}{(2{\pi}\hbar)^2}\delta(\varepsilon-\epsilon_{\bm p} + \mu), \label{normalDOS}
\end{align}
where $\epsilon_{\bm p}$ denotes the single particle energy.  
${p}_{x}$ and ${p}_{y}$ are the $x$ and $y$-components of the quasiparticle momentum, respectively, 
while $\mu$ is the chemical potential. 
We should also restrict momentum integration in Eq. (\ref{normalDOS}) to the first Brillouin zone. 
The superconducting DOS in the homogeneous system is  given by \cite{Ueki18}
%
\begin{align}
&N_{\rm s}(\varepsilon)
= N(0)\left\langle 
\frac{|\varepsilon|}{\sqrt{\varepsilon^2 - \Delta_{\rm bulk}^2 \phi^2}} 
\theta (|\varepsilon|- \Delta_{\rm bulk} |\phi |)
\right\rangle_{\rm F} \notag \\
& \ \ \ + N'(0)\left\langle  
{\rm sgn} (\varepsilon) \sqrt{\varepsilon^2 - \Delta_{\rm bulk}^2 \phi^2}
\theta (|\varepsilon|- \Delta_{\rm bulk} |\phi |)
\right\rangle_{\rm F}, \label{Ns}
\end{align}
where $\Delta_{\rm bulk}$ denotes the gap amplitude in the bulk.

The chemical potential difference 
between the normal and superconducting states of the homogeneous system
is given by\cite{Ueki18}
\begin{align}
\delta \mu = - \frac{N'(0)}{N(0)} \int_{\tilde{\varepsilon}_{{\rm c}-}}^{\tilde{\varepsilon}_{{\rm c}+}} d \varepsilon 
\varepsilon \bar{n}(\varepsilon) \left[\frac{N_{{\rm s}0}^{\rm bulk}(\varepsilon)}{N(0)} - 1 \right] - c \frac{N'(0)}{N(0)} \Delta_{\rm bulk}^2, \label{dmu}
\end{align}
where  $N_{{\rm s}0}^{\rm bulk}(\varepsilon)$ is the LDOS in the bulk obtained from the standard Eilenberger equations with the zeroth-order in $\delta$ as
\begin{align}
N_{{\rm s}0}^{\rm bulk}(\varepsilon) 
= N(0)\left\langle \frac{|\varepsilon|}{\sqrt{\varepsilon^2 - \Delta_{\rm bulk}^2 \phi^2}} 
\theta (|\varepsilon|- \Delta_{\rm bulk} |\phi |)\right\rangle_{\rm F}. 
\end{align}

The details of the derivation of Eqs. (\ref{LDOS}), (\ref{AQCEq}), (\ref{Ns}), and (\ref{dmu}) 
are available in Ref. \citen{Ueki18}.

\begin{figure}[t]
        \begin{center}
                \includegraphics[width=0.65\linewidth]{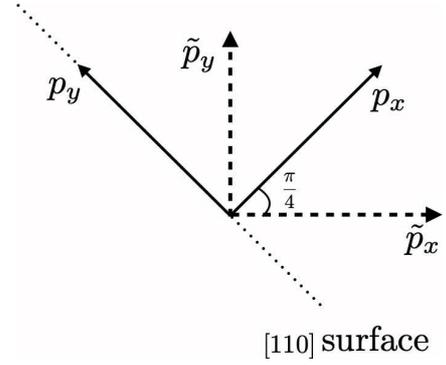}
                \end{center}
 \caption{Schematic representation of quasiparticle momentum axes $\tilde{p}_x$ and $\tilde{p}_y$ transformed to $p_x$ and $p_y$, respectively by a $\pi/4$ rotation. This rotation gives a [110] surface along the $p_y$ direction.}
\label{fig1.eps}
\end{figure}

\section{Numerical Results \label{sec:III}}
\subsection{Numerical procedures}
We here perform numerical calculations  for a quasi-two-dimensional semi-infinite system with a single specular surface at $x=0$. As a starting point, we introduce the single-particle energy on a two-dimensional square lattice 
used for high-$T_{\rm c}$ superconductors\cite{Kontani,KontaniText,Kita09},
%
\begin{align}
&\epsilon_{\bm p} 
= -2t\left(\cos \frac{\tilde{p}_xa}{\hbar}+\cos \frac{\tilde{p}_ya}{\hbar}\right) 
+4t_1\left(\cos \frac{\tilde{p}_xa}{\hbar} \cos \frac{\tilde{p}_ya}{\hbar}-1\right) \notag \\
&\ \ \ + 2t_2\left(\cos \frac{2\tilde{p}_xa}{\hbar} + \cos \frac{2\tilde{p}_ya}{\hbar}-2\right),\label{SPE}
\end{align}
with the lattice constant $a$,  
the dimensionless hopping parameters $t_1/t=1/6$ and $t_2/t=-1/5$, 
and the momenta $\tilde{p}_{x}$ and $\tilde{p}_{y}$ given by 
$\tilde{p}_{x}=(p_{x}+p_{y})/\sqrt{2}$ and $\tilde{p}_{y}=(p_{y}-p_{x})/\sqrt{2}$ as depicted in Fig. \ref{fig1.eps}. 
We also adopt a model $d$-wave pairing as $\phi=C(\cos \tilde{p}_{x} - \cos \tilde{p}_{y})$, 
and use ${\bm v}_{\rm s} = {\bm 0}$ in the absence of external magnetic field and without spontaneous edge currents. 
This is the same as using $\Delta = \Delta^*$ and ${\bm A} = {\bm 0}$. 
Here the real constant $C$ is determined 
via the normalization condition $\langle{\phi}^2\rangle_{\rm F}=1$, 
and ${\bm v}_{\rm s}$ is the superfluid density defined by ${\bm v}_{\rm s} \equiv (\hbar / m) ({\bm \nabla} \varphi - 2e{\bm A}/\hbar)$,
with $m$ and $\varphi$ denoting the electron mass and the phase of the pair potential, respectively.

First, we obtain the self-consistent solutions to the standard Eilenberger equations in Eqs. (\ref{Eeq}) 
and (\ref{gapeq}) 
using the Riccati method\cite{Nagato93,Schopohl95,Schopohl98,KitaText}.
%
The relevant boundary condition used in the bulk region is obtained by carrying out a gradient expansion\cite{KitaText} up to the first-order,
as shown in Appendix \ref{AppA}.
We also assume mirror reflection at the surface such that $Q(\varepsilon_n,{\bm p}_{\rm F},0) 
\equiv Q(\varepsilon_n,{\bm p}'_{\rm F},0)$, for an arbitrary function $Q$,
{where} ${\bm p}_{\rm F}$ and ${\bm p}'_{\rm F}$ are the Fermi momenta before 
and after reflection at the surface, respectively, and are related by 
${\bm p}'_{\rm F}$=${\bm p}_{\rm F}-2{\bm{n}}({\bm{n}}\cdot{{\bm p}_{\rm F}})$, 
with ${\bm{n}}$=$-\bm{\hat{x}}$\cite{Buchholtz95}.
We note that we need to solve the Riccati-type equation (see Eqs. (\ref{Riccatieq}) 
and (\ref{dpx_a}) in Appendix \ref{AppA}) by numerical integration 
towards the $-\hat{\bm x}$ direction for $v_{{\rm F}x}< 0$ 
from the bulk at $x = x_{\rm c} \gg \xi_0$ to the surface at $x=0$ 
and towards the $\hat{\bm x}$ direction for $v_{{\rm F}x} > 0$
from the surface at $x=0$ to the bulk at $x = x_{\rm c}$. 
We also use the solutions obtained by the gradient expansion of the Riccati-type equation
up to the first-order (see Appendix \ref{AppA}) in the region of $|v_{{\rm F}x}| \ll \langle v_{\rm F} \rangle_{\rm F}$, where $v_{\rm F}$ is the magnitude of the Fermi velocity.

Next, we solve Eq. (\ref{AQCEq}) to obtain the surface electric field 
with the boundary conditions
where the electric field vanishes at the surface  
and the first term on the LHS of Eq. (\ref{AQCEq}) is neglected near the bulk, 
using Eq. (\ref{g1Eq}) and substituting the Green's functions $f_0$ and $g^{\rm R}_{0}$ into Eq. (\ref{AQCEq}) accordingly.
We obtain the retarded Green's functions 
by performing the transformation $\varepsilon_n \to - i \varepsilon + \eta$
and using the same procedures as in the calculation of the Matsubara Green's functions.
The derivatives $\partial f_0 / \partial x$ and $\partial f_0 / \partial p_{{\rm F}x}$ in Eq. (\ref{g1Eq})  are also shown in Appendix A.
We then calculate the corresponding charge density using Gauss' law, 
${\bm \nabla}\cdot{\bm E(x)}=\rho(x)/d \epsilon_0$.
Furthermore, we calculate the LDOS, 
by substituting the retarded Green's functions $g^{\rm{R}}_{0,1}$ and $f^{\rm{R}}_{0}$ into Eq. (\ref{LDOS}).
We also use $g^{\rm R}_1=0$ in the bulk as a boundary condition to solve Eq. (\ref{g1Eq}).
We choose the parameters appropriate for cuprate superconductors
as $\delta = 0.05$, $t = 14 \Delta_0$, and $\lambda_{\rm TF} = 0.05 \xi_0$, 
where $\Delta_0$ denotes the gap amplitude at zero temperature.

\begin{figure}[t]
        \begin{center}
                \includegraphics[width=0.85\linewidth]{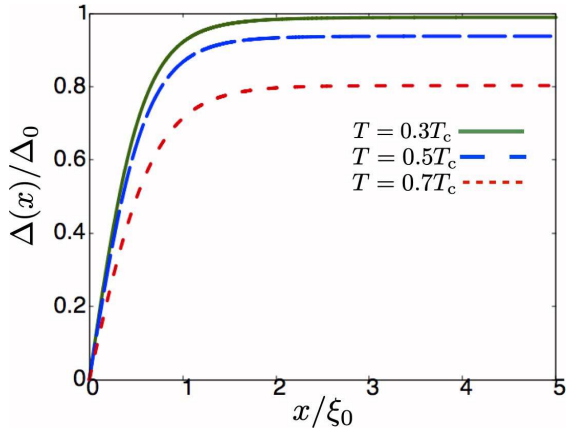}
                \end{center}
 \caption{(Color online) Temperature dependences of the self-consistent gap amplitude for the $d_{x^2-y^2}$-wave state with a smooth [110] surface at $x=0$. At temperatures $T=0.3T_{\rm c}$ (green solid line), $0.5T_{\rm c}$ (blue long dashed line), and $0.7T_{\rm c}$ (red short dashed line), for the filling $n=0.9$.}
\label{fig2}
\end{figure}

\begin{figure}

        \begin{center}
                \includegraphics[width=0.9\linewidth]{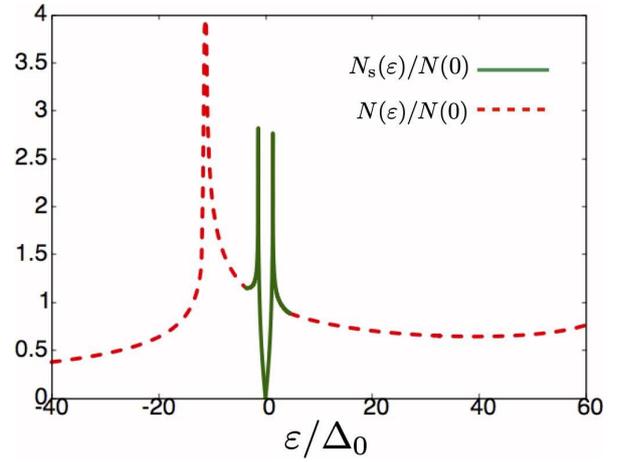}
                \end{center}
 \caption{(Color online) Superconducting DOS ${{N_{\rm s}}}(\varepsilon)$ (green solid line) 
 and the normal DOS $N(\varepsilon)$ (red dashed lines) in the homogeneous system 
 at temperature $T = 0.1 T_{\rm c}$, for the filling $n=0.9$, in units of $N(0)$ over $-40\Delta_0\leq\varepsilon\leq60\Delta_0$.}
\label{fig3}
\end{figure}

\begin{figure}
        \begin{center}
                \includegraphics[width=0.9\linewidth]{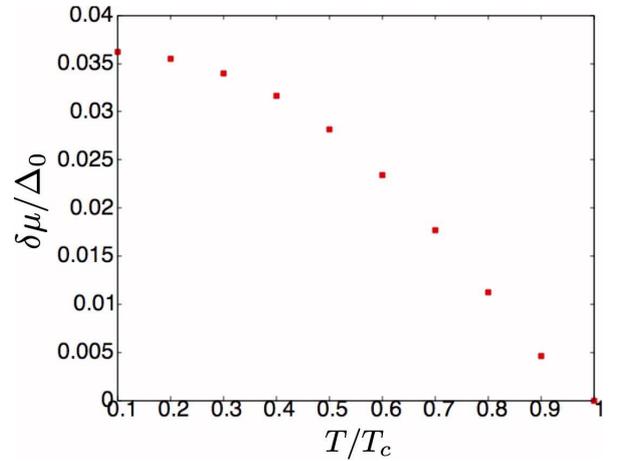}
                \end{center}
 \caption{(Color online) Temperature dependence of the chemical potential difference $\delta\mu$ between the normal and the superconducting states of the homogeneous system at the filling $n=0.9$.}
\label{fig4}
\end{figure}

\subsection{Results}
We discuss our numerical results in the following.
Figure \ref{fig2} shows the self-consistent gap amplitude 
for the $d$-wave paired superconductor with a [110] oriented surface
at the filling $n=0.9$. 
The pair potential shows spatial variation in the surface region.
The suppression of the pair-potential around the surface is caused by the symmetry of the pair potential given by $\phi(\bm{p}_{\rm{F}})=-\phi(\bm{p}'_{\rm{F}})$ and mirror reflection implies that $f_{0}(\varepsilon_n,{\bm p}_{\rm F},0) = f_{0}(\varepsilon_n,\bm {p}'_{\rm F},0)$\cite{Nagato95}.
The pair potential is approximately described by $\Delta(x)=\Delta_{\rm{bulk}}\tanh(x/\xi_{1})$, where $\xi_{1}$ is the healing length defined as $\xi_{1} = \lim_{x\to 0} x[\Delta_{\rm{bulk}}/\Delta(x)]$.
Note that $\xi_{1}$ is well described by the coherence length $\xi_{\rm{c}}\equiv [\braket{\hbar^{2}v^{2}_{{\rm{F}}x} \phi^{2} }_{\rm{F}}/\braket{\phi^{4} }_{\rm{F}}]^{\frac{1}{2}}\Delta^{-1}_{\rm{bulk}}$.

In Fig.\ \ref{fig3}, the superconducting DOS and the normal DOS in the homogeneous system 
at the filling $n=0.9$
connect at energies $\varepsilon = \tilde{\varepsilon}_{{\rm c}+}$ and $\tilde{\varepsilon}_{{\rm c}-}$.
They connect more smoothly taking into account higher order derivatives of the DOS 
at the Fermi level, although the higher-order derivatives contribute little to quantities. 
Thus, we can perform the following calculations using these cutoff energies $\tilde{\varepsilon}_{{\rm c}\pm}$. 
We also obtain $N'(0)\Delta_0/N(0) = -0.04819340$, $-0.03177848$, and $-0.01412786$ 
for $n=0.8$, $0.9$, and $1.15$, 
respectively, from the normal DOS calculations for each filling.
Here we note that the sign of $N'(0)\Delta_0/N(0)$ for all the fillings is negative, 
since the Van Hove singularity in the DOS obtained from Eq. (\ref{SPE}) is at about $n=0.5$, 
but not at a half filling. 
In Fig.\ \ref{fig4}, we show the temperature dependence of the chemical potential difference 
between the normal and superconducting states of the homogeneous system for the filling $n=0.9$
due to the SDOS at the Fermi level. 
Assuming roughly normal metal at the [110] surface, we can explain that the (effective) chemical potential difference between the surface and the bulk is what brings about the redistribution of charged particles.

\begin{figure}[t]
        \begin{center}
                \includegraphics[width=0.9\linewidth]{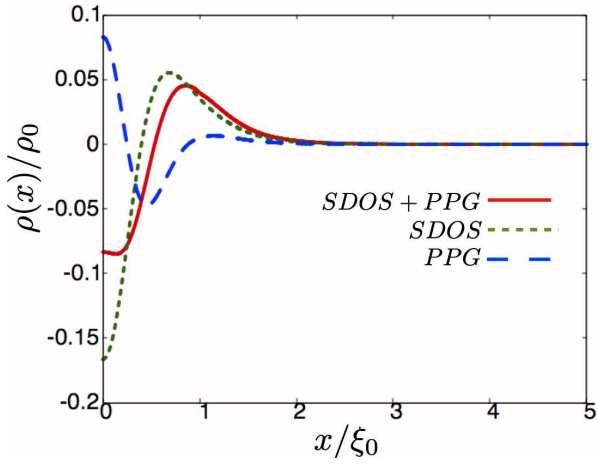}
                \end{center}
 \caption{(Color online) Total surface charge density (red solid line) due to the PPG force (blue long dashed line) and the SDOS pressure (green short dashed line), in units of $\rho_0\equiv d \epsilon_0 \Delta_0/|e|\xi_0^2$, at temperature $T=0.5T_{\rm{c}}$, for the filling $n=0.9$, with $\eta=0.01$.}
\label{fig5}
\end{figure}

\begin{figure}[t]
        \begin{center}
                \includegraphics[width=0.9\linewidth]{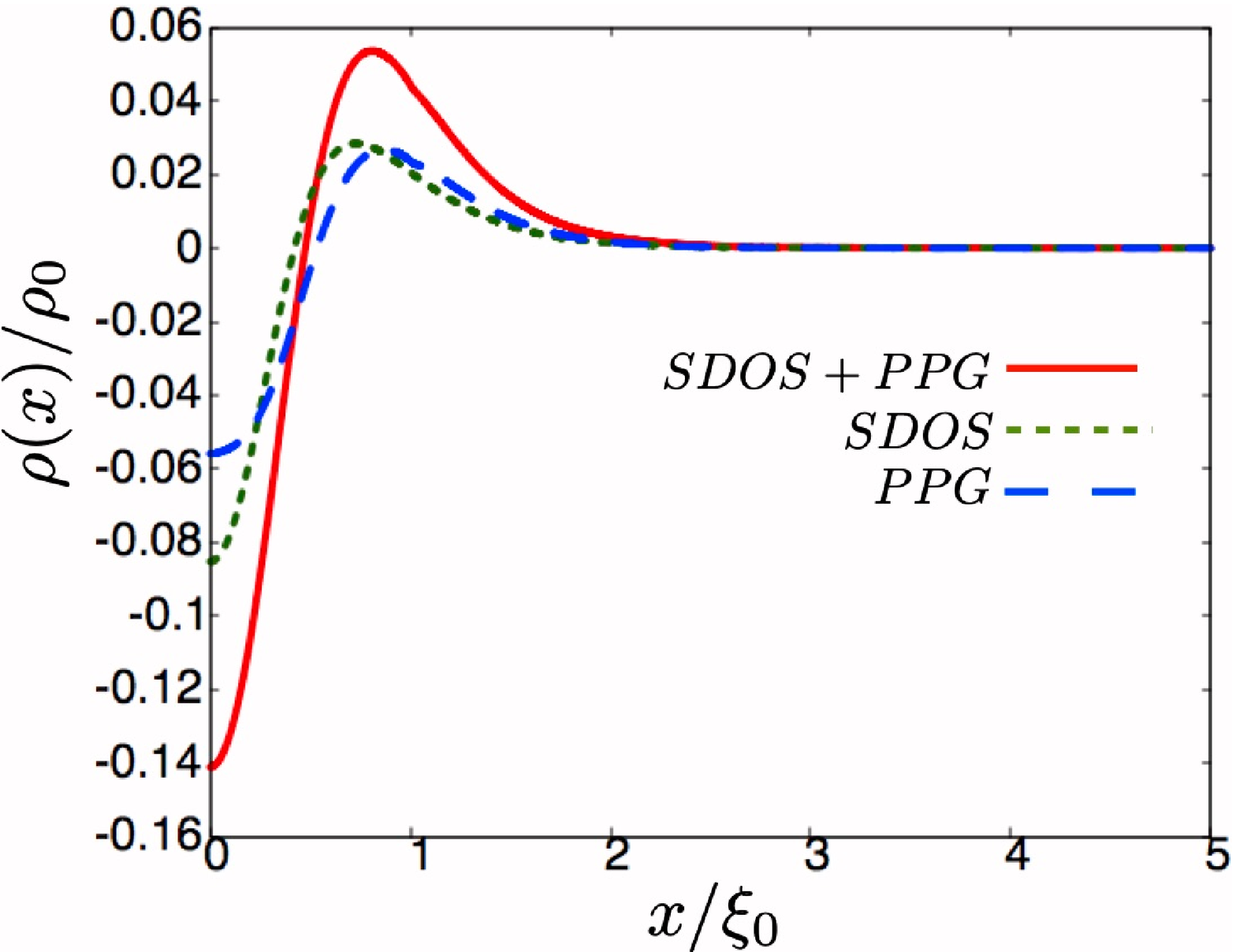}
                \end{center}
 \caption{(Color online) Total surface charge density (red solid line) due to the PPG force (blue long dashed line) and the SDOS pressure (green short dashed line), in units of $\rho_0\equiv d \epsilon_0 \Delta_0/|e|\xi_0^2$, at temperature $T=0.5T_{\rm{c}}$, for the filling $n=1.15$, with $\eta=0.01$.}
\label{fig6}
\end{figure}

Furthermore, Fig. \ref{fig5} shows the total surface charge  
with contributions from the PPG force and the SDOS pressure for the filling $n=0.9$.
The SDOS pressure gives the dominant contribution to the surface charge within our present model at $n=0.9$. 
We also confirmed that the SDOS pressure was dominant at not only $n=0.9$, 
but also at $n=0.8$ and $n=1.15$.
Since there is no supercurrents near the surface, 
there are no phase terms of the pair potential in the PPG force terms which are dominant in vortex systems\cite{Masaki19}. 
Therefore, the contribution from the PPG force to the surface charge becomes small, 
and the SDOS pressure is dominant in a wide parameter range of the semi-finite system, 
compared to the vortex system. 
Figure \ref{fig6} shows the total surface charge  
with contributions from the PPG force and the SDOS pressure for the filling $n=1.15$.
The PPG force contribution has the same negative sign as the SDOS pressure contribution, with the SDOS pressure giving 
the dominant contribution to the total charge. 
As shown in Fig.\ \ref{fig5} and \ref{fig6}, 
the sign of the charge density due to the PPG force at $n = 0.9$ is different from the one at $n = 1.15$. 
To explain this, we expand the quasiclassical Green's functions in terms of the pair potential up to third order, 
assume that the pair potential has the form $\Delta\simeq\Delta_{\rm bulk}\tanh(x/\xi_{1})$, and substitute them into Eq. (\ref{AQCEq}).
We then obtain the charge density due to the PPG force at $x=0$ as (see Appendix B for details)
\begin{align}
&\rho_{\rm{PPG}} (0) \sim  - \frac{2a^{(3)}\hbar^{2} d \epsilon_0 \Delta^{2}_{\rm bulk}}{\xi^{4}_{1}e}\left\langle{\phi^{2} \frac{{\partial} v_{{\rm{F}}x}}{{\partial} p_{{\rm{F}}x}}}\right\rangle_{\rm{F}} , \label{GL}
\end{align}
where $\rho_{\rm{PPG}}$ represents the charge density induced by the PPG force and $a^{(3)}\equiv \pi k_{\rm{B}}T \sum_{n=0}^{\infty}\varepsilon^{-3}_{n}$.
As described in Appendix B, this approximation is valid near the critical temperature.
According to Eq.\ (\ref{GL}), we see that the filling dependence is determined only by $\braket{\phi^{2} {\partial}v_{{\rm{F}}x}/{\partial}p_{{\rm{F}}x}}\simeq \braket{ {\partial}v_{{\rm{F}}x}/{\partial}p_{{\rm{F}}x}}\propto (\underline{R}^{(\rm{n})}_{{\rm{H}}})_{xx}= (\underline{R}^{(\rm{n})}_{{\rm{H}}})_{yy}$, where $\underline{R}^{(\rm{n})}_{\rm{H}}$ is the Hall coefficient \cite{Kita09}.
Therefore, the charge density induced by the PPG force also changes its sign around $n=1$ as shown in Fig.\ 1 of Ref.\ \citen{Kita09}, which is mainly caused by the change of Fermi-surface curvature. 
We here note that the filling at which the surface charge due to the PPG force changes its sign is different from that of the Van Hove singularity.

\begin{figure}[t]
        \begin{center}
                \includegraphics[width=0.9\linewidth]{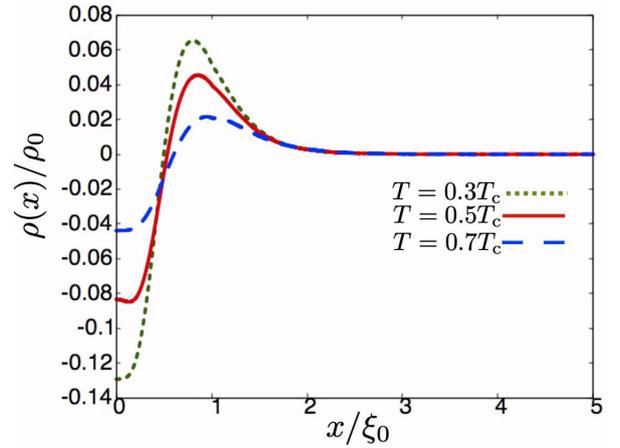}
                \end{center}
 \caption{(Colour online) Temperature dependence of the total surface charge induced by the PPG force and SDOS pressure, 
for the filling $n=0.9$, 
in units of $\rho_0\equiv d \epsilon_0 \Delta_0/|e|\xi_0^2$, with $\eta=0.01$, at temperatures $T=0.3T_{\rm c}$ (green short dashed line), $0.5T_{\rm c}$ (red solid line), and $0.7T_{\rm c}$ (blue long dashed line).}
\label{fig7}
\end{figure}

In Fig. \ref{fig7}, the temperature dependence of the total surface charge for the filling $n=0.9$ 
is shown. 
The total induced surface charge increases with a decrease in temperature.
This follows the temperature dependence of the slope of the pair potential shown in Fig. \ref{fig2}. 
One may notice that the temperature dependence of the second order derivative of $\rho(x\gtrsim0)$ $x$ is not monotonic.
The second order derivative of $\rho(x=0)$ is given by 
$({d}^2 \rho(x)/{d} x^{2})_{x=0}\propto -[\rho_{\rm{SDOS}}(0)/\xi^{2}_{\rm{SDOS}}+\rho_{\rm{PPG}}(0)/\xi^{2}_{\rm{PPG}}]$, 
where $\xi_{\rm{SDOS}}$ ($\xi_{\rm{PPG}}$) is defined by the value of $x$ at the first peak of the charge density due to the SDOS pressure (PPG force).
Thus, not only $\rho_{i}(0)$ but also $\xi_{i}$ is necessary when we consider $({d}^2 \rho(x)/{d} x^{2})_{x=0}$.
In the present case, although $\rho(0)$ decreases monotonically as the temperature decreases, $({d}^2 \rho(x)/{d} x^{2})_{x=0}$ behaves nonmonotonically because of the competition between $\rho_{\rm{SDOS}}(0)/\xi^{2}_{\rm{SDOS}}<0$ and $\rho_{\rm{PPG}}(0)/\xi^{2}_{\rm{PPG}}>0$.

\begin{figure}
        \begin{center}
                \includegraphics[width=0.9\linewidth]{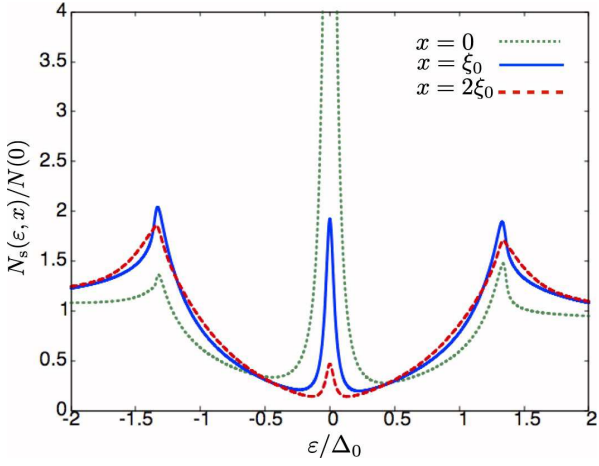}
                \end{center}
 \caption{(Color online) LDOS $N_{\rm s}(\varepsilon,x)$ at $x=0$ (green short dashes), $\xi_0$ (blue solid line), and $2\xi_0$ (red long dashes), with $\eta=0.04$, at temperature $T=0.1T_{\rm c}$, for the filling $n=0.9$.}
\label{fig8}
\end{figure}

\begin{figure}
        \begin{center}
                \includegraphics[width=0.9\linewidth]{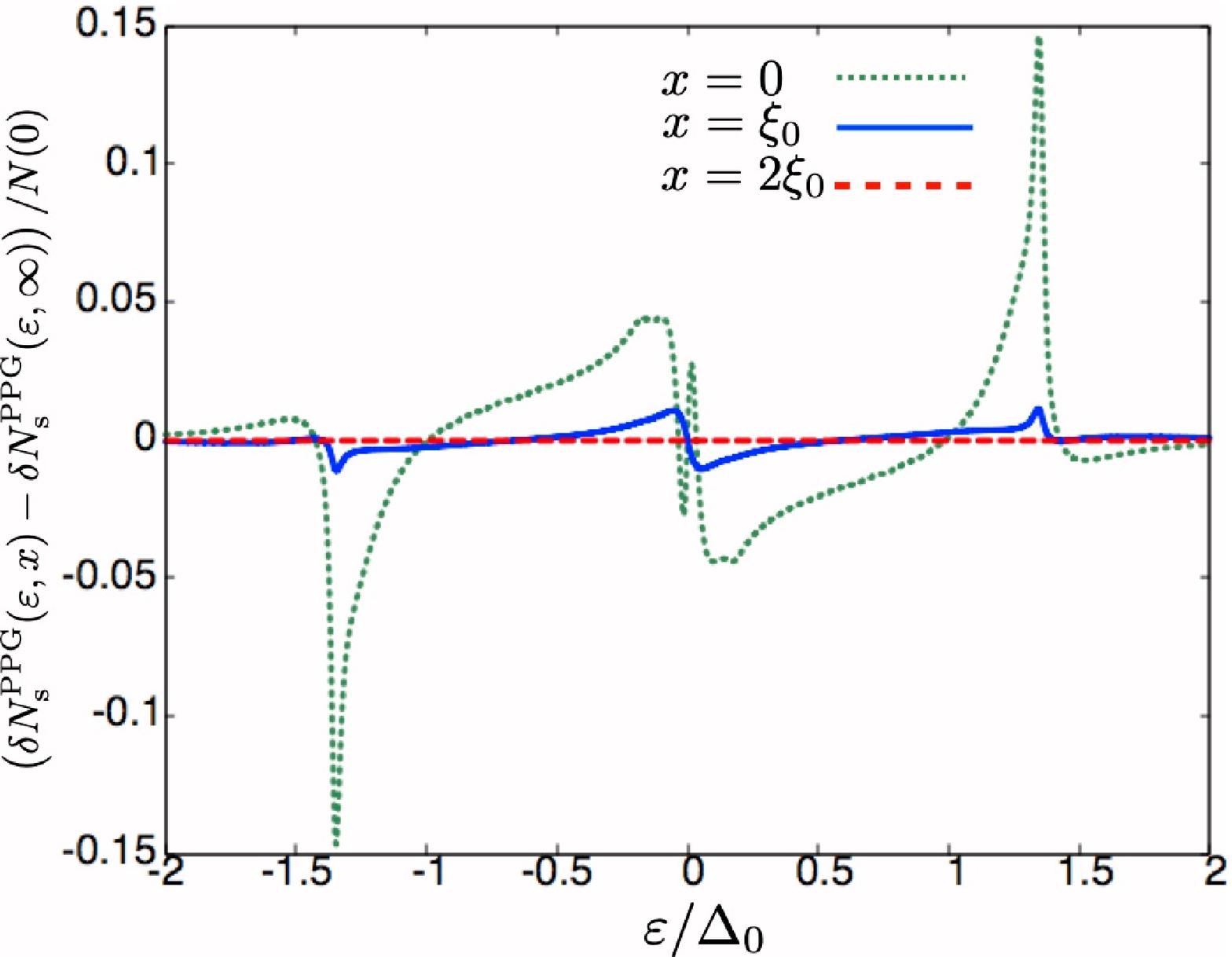}
                \end{center}
 \caption{(Color online) Deviation $ \delta N_{\rm s}^{\rm PPG}(\varepsilon, x)-\delta N_{\rm s}^{\rm PPG}(\varepsilon, \infty)$, where $\delta N_{\rm s}^{\rm PPG}(\varepsilon, x)$ is the deviation of the LDOS from the Eilenberger solution. We use $\eta=0.04$, temperature $T=0.1T_{\rm c}$, and at electron filling $n=0.9$.}
\label{fig9}
\end{figure}

\begin{figure}
\begin{center}
\includegraphics[width=0.9\linewidth]{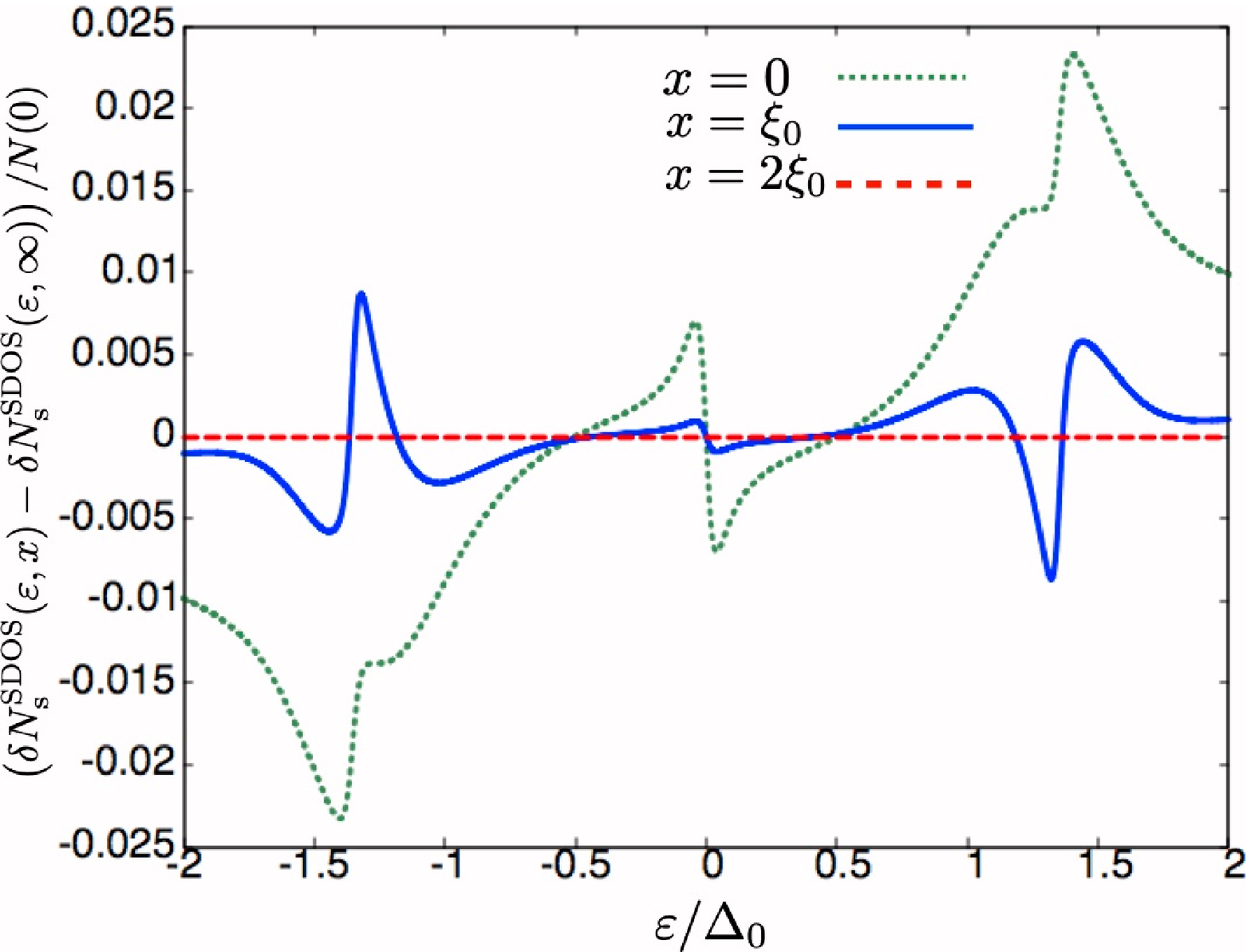}
\end{center}
\caption{
(Color online) Deviation $\delta N_{\rm s}^{\rm SDOS}(\varepsilon, x)- \delta N_{\rm s}^{\rm SDOS}(\varepsilon, \infty)$, where $\delta N_{\rm s}^{\rm SDOS}(\varepsilon, x)$ is the deviation of the LDOS from the Eilenberger solution. We use $\eta=0.04$, temperature $T=0.1T_{\rm c}$, and at electron filling $n=0.9$.
}
\label{fig10}
\end{figure}

\begin{figure}
        \begin{center}
                \includegraphics[width=0.9\linewidth]{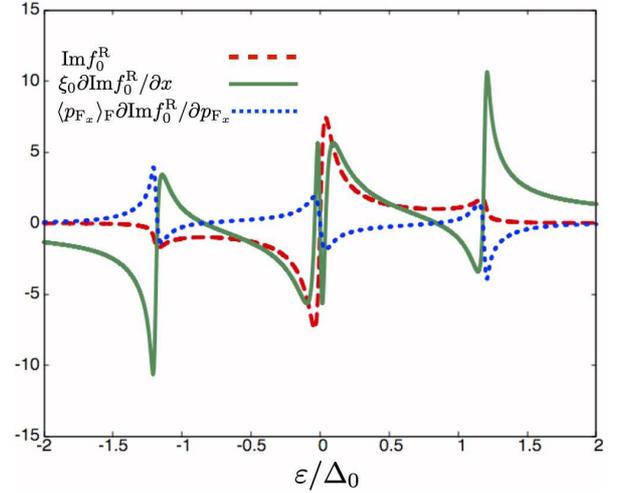}
                \end{center}
 \caption{(Color online) ${\rm Im}f_0^{\rm R}$, $\xi_0 \partial {\rm Im}f_0^{\rm R} / \partial x$, and $\langle p_{{\rm F}x} \rangle_{\rm F} \partial {\rm Im}f_0^{\rm R} / \partial p_{{\rm F}x}$ at the surface for the momentum direction $\varphi_{{\bm k}_{\rm F}}=\pi/6$ with $\eta =0.04$.}
\label{fig11}
\end{figure}

Emig {\it et al.} considered the surface charging in $d$-wave superconductors phenomenologically \cite{Emig}.
In fact, the right-hand side in Eq.\ (\ref{AQCEq}) reproduces the gradient of Eq.\ (3.1) in Ref.\ \citen{Emig} with the following approximations: (i) neglect the contributions from the PPG force, (ii) use the approximation for $\bm{\nabla}g^{\rm{R}}_{0} $ as $\bm{\nabla}g^{\rm{R}}_{0} \simeq \bm{\nabla}g^{(2)}_{0} (\varepsilon_{n}\to -i\varepsilon+\eta)$ (see Appendix B) (iii) $c\sim  \ln(\varepsilon_{\rm{c}}/k_{\rm{B}}T_{\rm{c}})$, where $\varepsilon_{\rm{c}}$ represents the energy cutoff of the order of the Debye temperature.
With these approximations, we obtain the electric-field equation equivalent to the gradient of Eq.\ (3.2) of Ref.\ \citen{Emig} as follows:
\begin{align} 
(-\lambda^{2}_{\rm{TF}}\nabla^{2}+1)\bm{E}_{\rm{SDOS}} \simeq  -\frac{\tilde{c}}{2e}\frac{N'(0)}{N(0)} \bm{\nabla}\Delta^{2}, \label{eleEmig}
\end{align}
where $\tilde{c} \equiv  \ln(\varepsilon_{\rm{c}}/k_{\rm{B}}T_{\rm{c}})$.
From this relation, we find the SDOS pressure is essentially equivalent to the mechanism studied in Ref.\ \citen{Emig}.
If we substitute $\Delta(x)\simeq \Delta_{\rm{bulk}} \tanh(x/\xi_{1})$ into Eq.\ (\ref{eleEmig}) with approximation $(-\lambda^{2}_{\rm{TF}}\nabla^{2}+1)\bm{E}_{\rm{SDOS}} \simeq \bm{E}_{\rm SDOS}$, we obtain charge density as
\begin{align} 
\rho_{\rm SDOS} \simeq  -\frac{\tilde{c}}{e}  \frac{N'(0) d \epsilon_0 \Delta^{2}_{\rm{bulk}}}{N(0)\xi^{2}_{1}} 
\Bigg[1- 4\tanh^{2}\Bigg(\frac{x}{\xi_{1}}\Bigg)+3\tanh^{4}\Bigg(\frac{x}{\xi_{1}}\Bigg)\Bigg],\label{chargeEmigexact}
\end{align}
which naturally satisfies the charge neutrality.
On the other hand, however, the charge density described in Eq.\ (3.4) of Ref.\ \citen{Emig} is derived by using the asymptotic approximation from the bulk to the surface for the source term of the electric-field equation, i.e., the behaviour of $\rho_{\rm{SDOS}}$ is quite different from the charge density by Ref.\ \citen{Emig} around the surface.
Specifically, the charge density by Emig {\em{et al.}} ($\rho_{\rm{E}}$) is described by
\begin{align} 
\rho_{\rm{E}}(x)=  -4\frac{\tilde{c}}{e} \frac{N'(0) d \epsilon_0 \Delta^{2}_{\rm bulk}}{N(0)\xi_{1}}\Bigg[\frac{e^{-x/\lambda_{\rm{TF}}}}{\lambda_{\rm{TF}}}-\frac{2e^{-2x/\xi_{1}}}{\xi_{1}}\Bigg],\label{chargeEmigapp}
\end{align}
Using Eqs.\ (\ref{chargeEmigexact}) and (\ref{chargeEmigapp}), we make a comparison of the accumulated charge per unit area within a surface layer of thickness $x_{\rm{c}}$ estimated from the phenomenological analysis by Emig {\it et al.} $Q_{\rm{p}}(x_{\rm{c}})$ and the charge from our microscopic calculation $Q_{\rm{m}}(x_{\rm{c}})$.
Defining their ratio as $\gamma(x_{\rm{c}})\equiv Q_{\rm{p}}/Q_{\rm{m}}$, we obtain 
\begin{align}
\gamma(x_{\rm{c}})&= \frac{\int_{0}^{x_{\rm{c}}} \rho_{\rm{E}}(x) dx}{\int_{0}^{x_{\rm{c}}} \rho_{\rm{SDOS}}(x) dx} \notag \\
&\sim -\frac{4\cosh^{2}(x_{\rm{c}}/\xi_{1}) (e^{-x_{\rm{c}}/\lambda_{\rm{TF}}}-e^{-2x_{\rm{c}}/\xi_{1}})}{\tanh(x_{\rm{c}}/\xi_{1})},\label{gamma}
\end{align}
where we use $\rho_{\rm SDOS}$ in Eq.\ (\ref{chargeEmigexact}) to derive the last expression.
From this quantity with $\lambda_{\rm{TF}}\ll\xi_{1}$ ($e^{-2\xi_{1}/\lambda_{\rm{TF}}}\sim 0$), we find $\gamma(x_{\rm{c}}=2\xi_{1})\sim 1.075 \sim 1 $, while $\gamma(x_{\rm{c}}=0)\sim 4\xi_{1}/\lambda_{\rm{TF}} \gg1$.
Note that in Ref.\ \citen{Emig} the accumulated charge near the surface with thickness $x_{\rm{c}}=\lambda_{\rm{TF}}\ln(\xi_{1}/2\lambda_{\rm{TF}})\sim\lambda_{\rm{TF}}$ was calculated.
Therefore, we conclude from this comparison that (1) our theory with approximations (i), (ii), (iii) reproduces the formalism used in Ref.\ \citen{Emig}, (2) our estimate of the accumulated charge within a surface layer of thickness $x_{\rm{c}}\gtrsim 2\xi_{1}$ is consistent with the estimate in Ref.\ \citen{Emig}.
 In addition, we emphasise the importance of the PPG-force contribution neglected in Ref.\ \citen{Emig}, since it competes with the SDOS pressure and yields nontrivial temperature dependence for the surface charge as shown in Fig.\ \ref{fig7}.

Figure \ref{fig8} plots the normalized LDOS for the filling $n=0.9$, 
at $x=0$, $\xi_0$, and $2\xi_0$. 
The zero energy peak structure appears as we move from the bulk towards the surface, 
and the particle-hole asymmetry exists in the LDOS even around the ZES 
although the peak of the bound states is at the zero energy 
when we choose $\eta$ to be small enough. 
This may seemingly contradict the symmetry consideration on the ZES at the [110] surface\cite{Yada,Sato}. 
However this particle-hole asymmetry around the zero energy exists 
even in the LDOS obtained from the calculation based on the BdG equations as Fig. 10(c) in Ref. \citen{Tanuma98} 
and based on the $T$-matrix method without the quasiclassical approximation
as Figs. 6 and 8 in Ref. \citen{Yada}, 
similar to our result on the surface LDOS. 
Therefore, it is possible that only non-zero energy states near the zero energy may contribute to the particle-hole asymmetry, but a detailed study based on the BdG equation may be needed to clarify that.

Figures \ref{fig9} and \ref{fig10} plot the difference in deviations
$\delta N_{\rm s}^{\rm PPG}(\varepsilon, {x})-\delta N_{\rm s}^{\rm PPG}(\varepsilon, {\infty})$ and $\delta N_{\rm s}^{\rm SDOS}(\varepsilon, {x})-\delta N_{\rm s}^{\rm SDOS}(\varepsilon, {\infty})$. Where $\delta N_{\rm s}^{\rm PPG(SDOS)}(\varepsilon,x )$ is the local deviation from the Eilenberger solution due to quantum corrections from the PPG force (SDOS pressure) at $x=0,\xi_0,2\xi_0$. While   
$\delta N_{\rm s}^{\rm PPG(SDOS)}(\varepsilon,x=\infty)$ is the local deviation from the Eilenberger solution due to quantum corrections from the PPG force (SDOS pressure) in the bulk region.
The deviations $\delta N_{\rm s}^{\rm PPG} (\varepsilon, x)$ and $\delta N_{\rm s}^{\rm SDOS} (\varepsilon, x)$
are defined by

\begin{subequations}
\begin{align}
\delta N_{\rm s}^{\rm PPG} (\varepsilon, x) = N(0) \langle {\rm Re} g_1^{\rm R} \rangle_{\rm F}, 
\end{align}
\begin{align}
\delta N_{\rm s}^{\rm SDOS} (\varepsilon, x) = N'(0) \varepsilon \langle {\rm Re} g_0^{\rm R} \rangle_{\rm F} 
+ \frac{N'(0)}{2} \Delta  \langle ({\rm Im} f_0^{\rm R} + {\rm Im} \bar{f}_0^{\rm R}) \phi \rangle_{\rm F}. 
\end{align}
\end{subequations}
We observe spatial variation in the local particle-hole asymmetry in the LDOS deviations.
The difference in LDOS deviations between the surface and the bulk due to the PPG force 
and due to the SDOS pressure show a change in their local peaks around $\varepsilon =0$ 
and around $\varepsilon = \pm 1.5\Delta_0$. 
Furthermore, the behavior of $\partial {\rm Im}f_0^{\rm R} / \partial x$ 
results in the multiple turning points in the LDOS deviation 
due to the PPG force around $\varepsilon=0$, 
since the PPG force and the LDOS terms are t
he space-momentum Moyal product for $\Delta \phi$ and ${\rm Im} f_0^{\rm R}$.
Figure \ref{fig11} plots ${\rm Im}f_0^{\rm R}$, $\partial {\rm Im}f_0^{\rm R} / \partial x$, 
and $\partial {\rm Im}f_0^{\rm R} / \partial p_{{\rm F}x}$ at the surface 
for the momentum direction $\varphi_{{\bm k}_{\rm F}}=\pi/6$ with $\eta =0.04$. 
Here $\varphi_{{\bm k}_{\rm F}}$ is defined by $\varphi_{{\bm k}_{\rm F}} \equiv \arctan k_{{\rm F}y}/k_{{\rm F}x}$, 
${\bm k}_{\rm F}$ is the Fermi momentum 
with the W point as the origin, 
and $k_{{\rm F}x}$ and $k_{{\rm F}y}$ have the same direction as $p_{{\rm F}x}$ and $p_{{\rm F}y}$. 
The spatial variation in the local electron-hole asymmetry suggests the presence of electric charging at the surface. 
More precisely, to obtain the variation of the charge density near the surface, the integration of the LDOS over negative energy range must be non-zero, 
since the expression for the charge density contains the distribution function\cite{Ueki18}. 
Although the LDOS deviation due to the PPG force is greater than that due to the SDOS pressure, 
its integral for the negative energy region is greater for the SDOS pressure than for the PPG force. 
Therefore, the surface charge due to the SDOS pressure is larger than that due to the PPG force. 
The connection between charging in superconductors and particle-hole asymmetry in the LDOS had already been discussed in the mixed state in type-II superconductors\cite{Masaki19,Hayashi}. On the other hand, unlike the vortex case, the contribution from the LDOS near the gap edge, rather than near zero energy, is dominant in the surface charge redistribution.

One may wonder why we can use the cutoff energies $\tilde{\varepsilon}_\pm$ obtained from the DOS in the homogeneous system for the calculation of the surface charge, 
despite the fact that the deviation in Fig. \ref{fig10} is large even at $\varepsilon = \pm 2 \Delta_0$. 
We again emphasize the following. Even if the cutoff energies are increased because the connection is not smooth, the difference between the LDOS only taking into account the first derivative of the DOS at the Fermi level and the normal DOS will only become larger. This does not satisfy Eq. (\ref{cutoffenergyeq}). To connect them more smoothly, we need to consider the higher-order derivative of the DOS at the Fermi level, but the higher-order derivatives contribute little to quantities. 
Therefore, we have used the same cutoff energies as those of the homogeneous system as an approximation.

\subsection{Proposed experiment}  
Surface charge measurement in superconductors may be possible using atomic force microscopy (AFM) in the non-contact mode. It is known that as the atomically sharp tip of the AFM cantilever approaches the surface of the superconductive material, the cloud of charged particles around the tip forms an electric dipole. The sample on the other hand, piles up charged particles in response to this dipole, so as to screen itself\cite{Peronio}. The observation of surface charge in this scenario reduces to observing the electrostatic interaction between the dipole at the tip of the cantilever and the charge at the surface of the sample.

\section{Summary \label{sec:IV}} 
In summary, we have performed a microscopic calculation on surface charging at a single [110] specularly reflective surface 
of a $d$-wave superconductor with a Fermi surface used for cuprate superconductors, 
using the augmented quasiclassical theory. 
We have shown that the SDOS pressure gives the dominant contribution to the charging compared to the PPG force, for all the realistic electron-fillings $n=0.8$, $0.9$ and $1.15$ at all temperatures. 
In addition, since the charge due to PPG force and that due to the SDOS pressure at $n=1.15$ have the same signs, the PPG force and the SDOS pressure may induce a larger surface charge in electron-doped $d$-wave superconductors compared to hole-doped superconductors.
Both the sign and amount of the surface charge depends greatly on the Fermi-surface curvature.
We have also calculated the LDOS within the augmented quasiclassical theory, 
taking into account the contributions due to the PPG force and the  SDOS pressure. 
At the surface, the LDOS shows a peak structure which signifies the presence of ZES. 
The bulk region shows a (nodal) gap-like structure which is a characteristic of the superconducting state. 
We have also shown the existence of particle-hole asymmetry 
(the SDOS pressure gives a locally larger particle-hole asymmetry at the filling $n=0.9$) in the LDOS. 
This spatially varying local asymmetry suggests the presence of electric charge.

Although our present study is restricted to a smooth surface without edge currents, 
the presence of surface roughness is expected to affect the surface states and may consequently alter the surface charge.
In addition, surface imperfections appear in the process of fabricating real samples, 
it is therefore important to consider the effects of surface imperfections, theoretically.
It is relatively easier to consider surface roughness within the quasiclassical theory 
using the random S-matrix theory\cite{Nagato96} or
by adding a disorder-induced self-energy\cite{Suzuki,Wang}.
Furthermore, in the presence of edge currents, the PPG force contribution to the charging effect may be enhanced due to the appearance of the phase terms of the pair potential \cite{Masaki19,Furusaki}. Other possible models for the $d$-wave surface state include the presence of a subdominant pairing near the surface and are characterized by spontaneously broken time-reversal symmetry\cite{Matsumoto,Fogelstrom98,Ryu,Matsubara}. Although the experimental identification of the evidence of these admixed states is still controversial, due to contradictory experimental data on cuprate superconductors\cite{Carmi,Salman}, they are very interesting. The effects of broken time-reversal symmetry on the charging properties at the surfaces of $d$-wave superconductors are not considered within our present model.
A combination of surface roughness and the presence of spontaneous edge currents may reveal very interesting physics  in relation to the surface charging in $d$-wave superconductors and chiral superconductors.

\begin{acknowledgments}
We are grateful to Marie Ohuchi for very insightful discussions. E. S. J. is supported by the Ministry of Education, Culture, Sports, Science, and Technology (MEXT) of Japan, H. U.  and W. K. are supported in part by JSPS KAKENHI Grant No. 15H05885 (J-Physics) and JSPS KAKENHI Grant Number 18J13241, respectively. The computation in this work was carried out using the facilities of the Supercomputer Center, 
the Institute for Solid State Physics, the University of Tokyo. 
\end{acknowledgments}

\appendix 
\section{Boundary conditions based on gradient expansion \label{AppA}}
We start from the Riccati form of Eq. (\ref{Eeq})\cite{Nagato93,Schopohl95,Schopohl98,KitaText}.
\begin{align}
v_{{\rm F}x}\frac{\partial{a}}{\partial{x}}=
-2\varepsilon_na-\Delta\phi{a^2}+\Delta\phi, \label{Riccatieq}
\end{align}
where $a$=$a(\varepsilon_n,{\bm p}_{\rm F},x)$ is the Riccati parameter 
and is related to $f_{0}$ and $g_{0}$ 
as 
\begin{align}
f_0 = \frac{2a}{1+a\bar{a}}, \ \ \ g_0=\frac{1-a\bar{a}}{1+a\bar{a}}.
\end{align}
We carry out a gradient expansion\cite{KitaText} of Eq. (\ref{Eeq}) using the expansion $a \approx a^{(0)} + a^{(1)}$,
which gives
\begin{align}
a^{(0)}& = \frac{\Delta \phi}{\varepsilon_n + \sqrt{\varepsilon_n^2 + \Delta^2 \phi^2}}, \notag \\ 
a^{(1)} &= - \frac{v_{{\rm F}x}}{2 \sqrt{\varepsilon_n^2 + \Delta^2 \phi^2}} \frac{\partial a^{(0)}}{\partial x}, \notag \\
\frac{\partial a^{(0)}}{\partial x} &= - \frac{a^{(0)}{}^2}{\sqrt{\varepsilon_n^2 + \Delta^2 \phi^2}}
\frac{d \Delta}{d x} \phi 
+\frac{a^{(0)}}{\Delta} \frac{d \Delta}{d x}. 
\end{align}

The derivatives $\partial f_0 / \partial x$ and $\partial f_0 / \partial p_{{\rm F}x}$ in Eq. (\ref{g1Eq}) are expressed as

\begin{align}
\frac{\partial f_0}{\partial x} &= \frac{2}{(1+a \bar{a})^2} \left( \frac{\partial a}{\partial x} - a^2 \frac{\partial \bar{a}}{\partial x} \right), \notag \\ 
\frac{\partial f_0}{\partial p_{{\rm F}x}} &= \frac{2}{(1+a \bar{a})^2} \left( \frac{\partial a}{\partial p_{{\rm F}x}} - a^2 \frac{\partial \bar{a}}{\partial p_{{\rm F}x}} \right). 
\end{align}
$\partial a / \partial x$ is obtained from Eq. (\ref{Riccatieq}), and $\partial a / \partial p_{{\rm F}x}$ is given
by solving the following equation:  
\begin{align}
&v_{{\rm F}x} \frac{\partial }{\partial x} \frac{\partial a}{\partial p_{{\rm F}x}}
= -2 \varepsilon_n \frac{\partial a}{\partial p_{{\rm F}x}} 
- \Delta \frac{\partial \phi}{\partial p_{{\rm F}x}} a^2 \notag \\
& \ \ \ -2 \Delta \phi a \frac{\partial a}{\partial p_{{\rm F}x}}
+ \Delta \frac{\partial \phi}{\partial p_{{\rm F}x}}
- \frac{\partial v_{{\rm F}x}}{\partial p_{{\rm F}x}} \frac{\partial a}{\partial x}, \label{dpx_a}
\end{align}
where the boundary condition for Eq. (\ref{dpx_a}) used near the bulk is given by
\begin{subequations}
\begin{align}
&\frac{\partial a}{\partial p_{{\rm F}x}} \approx \frac{\partial a^{(0)}}{\partial p_{{\rm F}x}} 
+ \frac{\partial a^{(1)}}{\partial p_{{\rm F}x}}, \notag \\
&\frac{\partial a^{(0)}}{\partial p_{{\rm F}x}} = 
- \frac{a^{(0)}{}^2}{\sqrt{\varepsilon_n^2 + \Delta^2 \phi^2}}
\Delta \frac{\partial \phi}{\partial p_{{\rm F}x}}
+\frac{a^{(0)}}{\phi} \frac{\partial \phi}{\partial p_{{\rm F}x}}, \notag \\
&\frac{\partial a^{(1)}}{\partial p_{{\rm F}x}} = 
\frac{v_{{\rm F}x} \Delta^2 \phi}{4 (\varepsilon_n^2 + \Delta^2 \phi^2)^2} 
\frac{\partial \phi}{\partial p_{{\rm F}x}} \frac{\partial a^{(0)}}{\partial x} \notag \\
& \ \ \ - \frac{1}{2\sqrt{\varepsilon_n^2 + \Delta^2 \phi^2}} 
\left( \frac{\partial v_{{\rm F}x}}{\partial p_{{\rm F}x}} \frac{\partial a^{(0)}}{\partial x} 
+ v_{{\rm F}x} \frac{\partial^2 a^{(0)}}{\partial x \partial p_{{\rm F}x}} \right), \notag \\
&\frac{\partial^2 a^{(0)}}{\partial x \partial p_{{\rm F}x}} = 
\frac{a^{(0)}{}^2}{(\varepsilon_n^2 + \Delta^2 \phi^2)^{3/2}} 
\Delta^2 \phi^2 \frac{d \Delta}{d x} \frac{\partial \phi}{\partial p_{{\rm F}x}} \notag \\
& \ \ \ - 2 \frac{a^{(0)}}{\sqrt{\varepsilon_n^2 + \Delta^2 \phi^2}} 
\frac{\partial a^{(0)}}{\partial x} \Delta \frac{\partial \phi}{\partial p_{{\rm F}x}} \notag \\
& \ \ \ - \frac{a^{(0)}{}^2}{\sqrt{\varepsilon_n^2 + \Delta^2 \phi^2}} 
\frac{d \Delta}{d x} \frac{\partial \phi}{\partial p_{{\rm F}x}}
+ \frac{\partial a^{(0)}}{\partial x} \frac{1}{\phi} \frac{\partial \phi}{\partial p_{{\rm F}x}}.
\end{align}
\end{subequations}

\section{Derivation of Eq.\ (\ref{GL}) \label{App}}
Here, we derive Eq.\ (\ref{GL})
using the following steps: (i) expand the quasiclassical Green's functions in terms of pair potential \cite{KitaText} up to third order based on the assumptions $|\Delta(x)|\ll \Delta_{0}$ and $\hbar v_{{\rm{F}}x}(d^{n} \Delta/d x^{n})=O(\Delta^{n+1})$, which are valid near $T_{\rm{c}}$, (ii) substitute the expanded Green's functions into the first term of Eq.\ (\ref{AQCEq}), and (iii) neglect the Thomas--Fermi term of Eq.\ (7).
Here, we consider the solutions only in $\varepsilon_{n}>0$.

First, we expand Green's functions as
\begin{align}
f_{0}=\sum_{\nu=1}^{\infty} f^{(\nu)}_{0}, \ g_{0}=1+\sum_{\nu=2}^{\infty} g^{(\nu)}_{0}. \label{Expand}
\end{align}
Substituting Eq.\ (\ref{Expand}) into Eq.\ (\ref{Eeq}) with the initial conditions $f^{(0)}_{0}=0$, $g^{(0)}_{0}=1$, and $g^{(1)}_{0}=0$, we obtain the following recursive relation:
\begin{align}
 f^{(\nu)}_{0}=\Bigg[\frac{\Delta\phi g^{(\nu-1)}_{0}}{\varepsilon_{n}}-\frac{\hbar  v_{{\rm{F}}x} }{2\varepsilon_{n}}\frac{\partial f^{(\nu-1)}_{0}}{\partial x}\Bigg] ,\label{fnugnu}
\end{align}
where we assumed $\hbar v_{{\rm{F}}x}d \Delta/d x=O(\Delta^{2})$.
Using Eq.\ (\ref{fnugnu}) with initial conditions and normalization condition $g^{2}_0= 1-f_0\bar{f}_0$, $f^{(\nu)}_{0}$ and $g^{(\nu)}_{0}$ up to third order are derived as follows
\begin{subequations}
\begin{align}
& f^{(1)}_{0}= \frac{\Delta\phi}{\varepsilon_{n}}, \ f^{(2)}_{0}= -\frac{\hbar \phi v_{{\rm{F}}x}  }{2\varepsilon^{2}_{n}}\frac{d \Delta}{d x},\notag\\
& f^{(3)}_{0}=-\Bigg[\frac{(\Delta\phi)^{2}}{2\varepsilon^{3}_{n}}-\frac{(\hbar  v_{{\rm{F}}x}  )^{2}}{4 \varepsilon^{3}_{n} }\frac{d^{2}}{d x^{2}}\Bigg]\Delta\phi,\\
&g^{(2)}_{0}= -\frac{(\Delta\phi)^{2}}{2\varepsilon^{2}_{n}}, \ g^{(3)}_{0}=0. 
  \end{align}
\end{subequations}
where we used ${\rm{Im}}\Delta=0$.

The equation for $\partial {\rm{Im}}g_{1}/\partial x$, which is included in the electric field equation in Eq.\ (\ref{AQCEq}), is given by
\begin{align}
 \frac{\partial {\rm{ Im}}g_{1}}{\partial x} 
&= \frac{-\hbar^{2}}{4\varepsilon^{3}_{n}} \Bigg[\Bigg(2  \phi^{2}\frac{\partial v_{{\rm{F}}x}}{\partial p_{{\rm{F}}x}}+ v_{{\rm{F}}x} \phi\frac{\partial \phi}{\partial p_{{\rm{F}}x}}\Bigg) \frac{d \Delta}{d x}\frac{d^{2} \Delta}{d x ^{2}}\notag\\
&-v _{{\rm{F}}x} \phi \frac{\partial  \phi}{\partial  p_{{\rm{F}} x}} \Delta\frac{d^{3} \Delta}{d x ^{3 }} \Bigg].\label{withoutmag}
\end{align}
If we only consider the PPG force, the electric field equation is given by the first term of Eq.\ (\ref{AQCEq}).
Thus, within the present approximation, the electric field due to the PPG force is given by
 \begin{align}
&\Big(-\lambda^{2}_{\rm{TF}}\frac{d^{2}}{d x^{2}}+1\Big)E_{{\rm{PPG}}x}(x)\notag\\
&\simeq  \frac{\hbar^{2} a^{(3)}}{2e } 
\Bigg[\Bigg(2\Big\langle\phi^{2}\frac{\partial v_{{\rm{F}}x}}{\partial p_{{\rm{F}}x}} \Big\rangle_{\rm{F}}+\Big\langle v_{{\rm{F}}x}\phi\frac{\partial \phi}{\partial p_{{\rm{F}}x}} \Big\rangle_{\rm{F}}\Bigg) \frac{d \Delta }{d x}\frac{d^{2} \Delta}{d x ^{2}}\notag\\
&-\Big\langle v_{{\rm{F}}x}\phi\frac{\partial \phi}{\partial p_{{\rm{F}}x}} \Big\rangle_{\rm{F}} \Delta \frac{d^{3} \Delta}{d x ^{3}}\Bigg],
\end{align}
where $E_{{\rm{PPG}}x}$ represents the $x$-component of electric field induced by the PPG force $a^{(3)}\equiv  \pi k_{\rm{B}}T \sum_{ 0 \leq  n\leq n_{\rm{c}}}\varepsilon^{-3}_{n}$.
Moreover, neglecting the term related to the Thomas--Fermi screening length, and using the assumption $\lambda_{\rm{TF}}\ll \xi_{0}$, we obtain the approximated charge density as
\begin{align}
& \rho_{\rm{PPG}} (x)\simeq  \frac{\hbar^{2} a^{(3)} \epsilon_{0}}{2e } 
\Bigg[\Bigg(2\Big\langle\phi^{2}\frac{\partial v_{{\rm{F}}x}}{\partial p_{{\rm{F}}x}} \Big\rangle_{\rm{F}}+\Big\langle v_{{\rm{F}}x}\phi\frac{\partial \phi}{\partial p_{{\rm{F}}x}} \Big\rangle_{\rm{F}}\Bigg)
\Bigg(\frac{d^{2} \Delta }{d x^{2}}\Bigg)^{2}\notag\\
&+ 2\Big\langle\phi^{2}\frac{\partial v_{{\rm{F}}x}}{\partial p_{{\rm{F}}x}} \Big\rangle_{\rm{F}}\frac{d \Delta }{d x}\frac{d^{3} \Delta}{d x ^{3}}
-\Big\langle v_{{\rm{F}}x}\phi\frac{\partial \phi}{\partial p_{{\rm{F}}x}} \Big\rangle_{\rm{F}} \Delta \frac{d^{4} \Delta}{d x ^{4}}\Bigg]. \label{chargeGL}
\end{align}
where $\rho_{\rm{PPG}} (x)\equiv d \epsilon_{0}(d E_{{\rm{PPG}}x}/d x)$.
Therefore, $\rho_{\rm{PPG}}(x)$ is expressible only in terms of $d^{n} \Delta(x)/d x^{n}$ in high-temperature region where the present approximation is valid.
Substituting the pair potential assumed as $\Delta(x)\simeq \Delta_{\rm bulk}\tanh (x/\xi_{1})$ into Eq.\ (\ref{chargeGL}) and taking the limit $x\to0$, we arrive at Eq.\ (\ref{GL}).

\end{document}